\documentclass[11pt]{article}

\usepackage{amsmath,amssymb,color,verbatim}
\usepackage{pstricks}
\usepackage[margin=2cm]{geometry}
\usepackage{graphicx}
\usepackage{framed}

\newcommand{\bnabla}{\boldsymbol{\nabla}}

\newcommand{\varep}{\varepsilon}
\newcommand{\pt}{\partial_{t}}
\newcommand{\bV}{\mathbi{V}}

\newcommand\shalf{\ensuremath{{\scriptstyle\frac{1}{2}}}}

\newcommand\sthird{\ensuremath{{\scriptstyle\frac{1}{3}}}}

\newcommand{\rem}[1]{}
\newtheorem{theorem}{Theorem}

\DeclareMathAlphabet{\mathbi}{OML}{cmm}{b}{it} 
\newcommand{\bx}{\mathbi{x}}
\newcommand{\bn}{\mbox{\boldmath$\hat{n}$}}
\newcommand{\bel}{\begin{equation}\label}
\newcommand{\ee}{\end{equation}}
\newcommand{\ben}{\begin{enumerate}}
\newcommand{\een}{\end{enumerate}}
\newcommand{\bde}{\begin{description}}
\newcommand{\ede}{\end{description}}
\newcommand{\bit}{\begin{itemize}}
\newcommand{\eit}{\end{itemize}}
\newcommand{\bc}{\begin{center}}
\newcommand{\ec}{\end{center}}

\newcommand{\bdF}{\mbox{\boldmath$F$}}
\newcommand{\bdf}{\mbox{\boldmath$f$}}

\newcommand{\bA}{\mbox{\boldmath$\mathcal{A}$}}

\newcommand{\bdB}{\mbox{\boldmath$\mathcal{B}$}}
\newcommand{\bdD}{\mbox{\boldmath$\mathcal{D}$}}

\newcommand{\bu}{\mathbi{u}}
\newcommand{\bv}{\mathbi{v}}
\newcommand{\bS}{\mathbi{S}}
\newcommand{\bU}{\mbox{\boldmath$\mathcal{U}$}}
\newcommand{\bcalJ}{\mbox{\boldmath$\mathcal{J}$}}
\newcommand{\bsfU}{\textsf{\textbf{U}}}
\newcommand{\bsfB}{\textsf{\textbf{B}}}
\newcommand{\bsfD}{\textsf{\textbf{D}}}
\newcommand{\sfS}{\textsf{S}}
\newcommand{\sfq}{\textsf{q}}
\newcommand{\sfQ}{\textsf{Q}}

\newcommand{\bom}{\mbox{\boldmath$\omega$}}

\newcommand{\bzeta}{\mbox{\boldmath$\zeta$}}

\newcommand{\bk}{\mbox{\boldmath$\hat{k}$}}

\newcommand{\beq}{\begin{eqnarray}\label} 
\newcommand{\eeq}{\end{eqnarray}}

\newcommand{\non}{\nonumber}
\newcommand{\Rey}{Re}

\makeatletter
\@addtoreset{equation}{section}

\makeatother

\begin{document}
\sf
\bc
\textbf{\Large The dynamics of the gradient of potential vorticity}
\par\vspace{7mm}
\textbf{\large J. D. Gibbon and D. D. Holm}
\par\vspace{1mm}
Department of Mathematics, Imperial College London SW7 2AZ, UK
\par\vspace{1mm}
{\small email: j.d.gibbon@ic.ac.uk and d.holm@ic.ac.uk}
\ec

\vspace{1mm}

\begin{abstract}
The transport of the potential vorticity gradient $\bnabla{q}$ along surfaces of constant 
potential temperature $\theta$ is investigated for the stratified Euler, Navier-Stokes and 
hydrostatic primitive equations of the oceans and atmosphere, in terms of the divergenceless flux 
vector $\bdB = \bnabla Q(q)\times\bnabla\theta$, for any smooth function $Q$ of of the potential 
vorticity $q$. The flux vector $\bdB$ is shown to satisfy a transport equation
reminiscent of that for magnetic field flux in magnetohydrodynamics.\rem{ 
$$
\partial_t\bdB - \mbox{curl}\,(\bU\times\bdB) = - 
\bnabla\big[qQ'(q)\,\mbox{div}\,\bU\big]\times\bnabla\theta\,,
$$
where $\bU$ is a formal transport velocity. While the left hand side of this expression 
is, the non-zero right 
hand side means that $\bdB$ is not frozen into the flow of $\bU$ when $\mbox{div}\,\bU \neq 0$. 
}
The result may apply to satellite observations of potential vorticity and potential temperature 
at the tropopause.
\end{abstract}
\par\vspace{10mm}
\bc
14th March 2010
\ec

\newpage
\section{\sf\large\textbf{Introduction}}

Potential vorticity (PV) is believed to be a particularly significant quantity in the dynamics of 
the atmosphere 
and the oceans \cite{Hos82,HMR85}. For an incompressible fluid the PV density is defined as $q = \bom
\cdot\bnabla\theta$ where $\bom = \mbox{curl}\,\bu$ is the vorticity for a fluid with a divergence-less
velocity field $\bu$ and $\theta$ is the potential temperature. This paper exposes a mechanism for 
creating the large gradients in potential vorticity density $\bnabla q$. Its discussion is based on 
the geometric properties of transport of intersections of level sets of the quantities $q$ and $\theta$ 
encoded in the vector $\bdB = \bnabla Q(q)\times\bnabla\theta$ which turns out to obey an evolution 
equation of the form\footnote{\sf \scriptsize The vector $\bdB = \bnabla Q(q)\times\bnabla\theta$ 
{has been discussed in \cite{KurgTat87,KurgPis00,Kurg02}} and should not be confused with 
the cross product $\bnabla B\times\bnabla\theta$ considered in \cite{Schaer93}, in which the scalar 
$B$ is a steady Bernoulli function.} 
\bel{froz1}
\partial_t\bdB - \mbox{curl}\,(\bU\times\bdB) = \bdD\,.
\ee
Here, the vector field $\bU$ is formally a transport velocity and will be derived explicitly below 
in several different cases. The quantity $\bdD$ expresses the rate of change of the flux of $\bdB$ 
in a frame moving with velocity  $\bU$. When $\bdD=0$, the flux of $\bdB$ is said to be \emph{frozen} 
into the flow with velocity  $\bU = \bu$. The Euler equations for the stretching and folding of vorticity, 
and the ideal MHD equations for the frozen-in transport of the flux of divergence-free magnetic field, 
both take the form (\ref{froz1}) with $\bdD=0$ {(see \cite{KurgTat87,KurgPis00,Kurg02})}. 
{However, the vector $\bdB$ turns out to be a wise choice even with dissipation\,: the 
vorticity dynamics of the Navier-Stokes equations with viscosity and the magnetic field evolution in 
MHD with resistivity, both take the form of equation (\ref{froz1}) but, because $\bdD\ne0$, the 
corresponding fields are no longer frozen into the flow.}
Figure 1 shows (model assimilated) satellite data for contours of potential temperature $\theta$ on the 
constant level surface of potential vorticity q = 2, which lies near the tropopause. See \cite{ECMWF} 
for animations of this data in frames taken every six hours. In these animations, the evolution of the 
contours of  $\theta$ on a level set of potential vorticity $q=2$ is seen\,: its appearance suggests 
the \emph{stirring} of one liquid in another by stretching and folding, {such as cream 
in black coffee.}

\begin{figure}[h]
\bc
\rotatebox{0}{\includegraphics[width=0.6\textwidth]{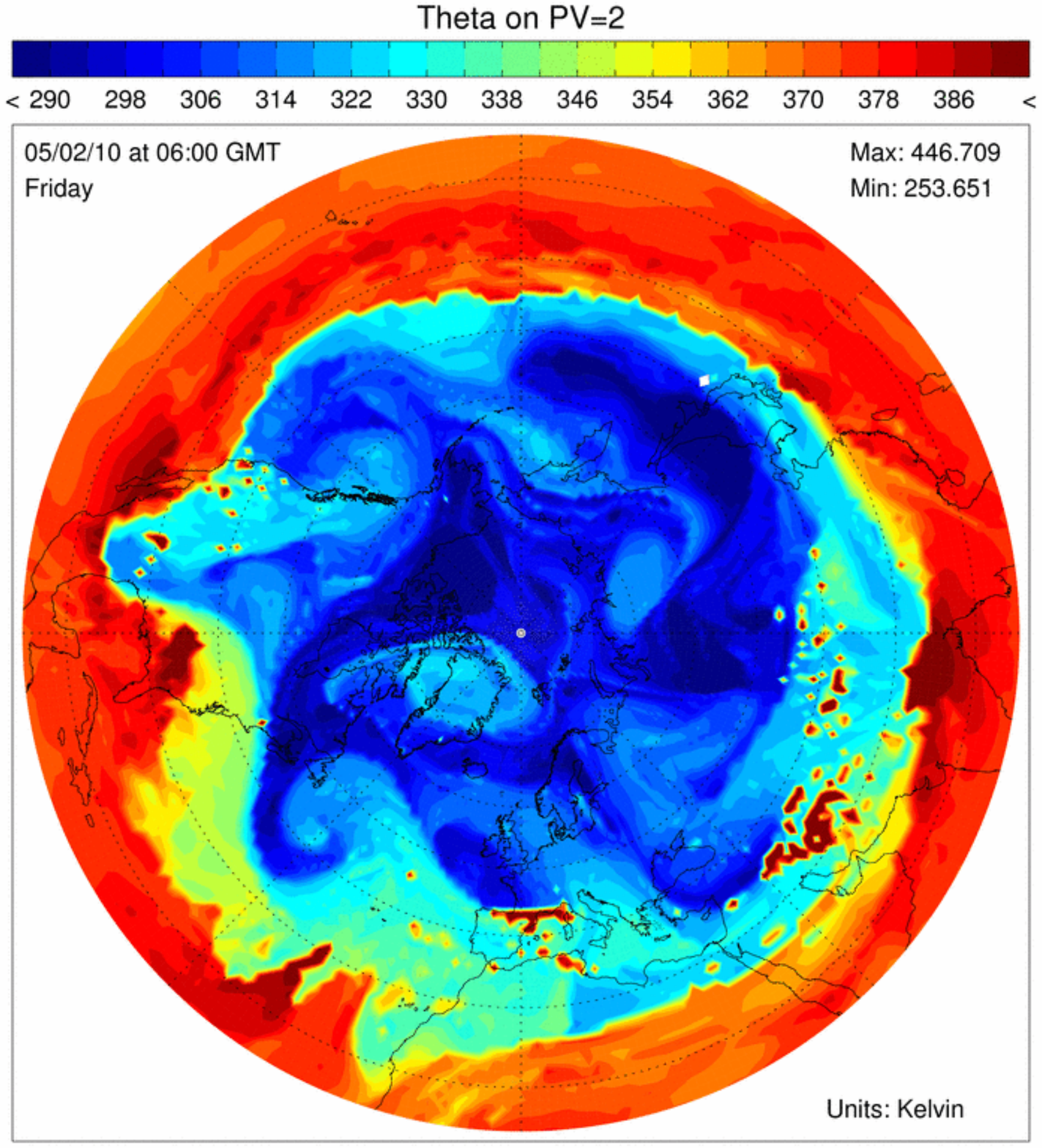}}
\caption{\sf\scriptsize A typical snapshot of satellite data taken every six hours of 
the Northern Hemisphere from ECMWF \cite{ECMWF} shows contours of potential temperature 
$\theta$ on a level set of potential vorticity $q=2$ located near the tropopause. Notice 
that high gradients of $\theta$ lie at the sharp interfaces of the contours. \label{Flo} }
\ec
\end{figure}

The object of the present work is to derive exact equations of the form (\ref{froz1}) for the 
evolution of the intersections of level sets of $q$ and $\theta$, in which we will find that 
$\bdD$ is given by the divergence-less vector 
\bel{froz2}
\bdD = - \bnabla\big[qQ'(q)\,\mbox{div}\,\bU\big]
\times\bnabla\theta
\,,
\ee
for any choice of the smooth function $Q$, and the vector $\bU$ will be derived in several cases. 
\smallskip

Equations (\ref{froz1}) and (\ref{froz2}) are first derived in \S2 in the narrower setting of the 
Euler and Navier-Stokes equations.  In \S3 the corresponding results are derived for the viscous 
hydrostatic primitive equations (HPE), which are commonly used in numerical simulations of the 
weather, climate and oceans.  The stretching and folding mechanism inherent in (\ref{froz1}) 
in the context of the viscous HPE may have some application to potentially rapid growth of 
$\bnabla q$ in the atmosphere and oceans, where the occurrence of extreme events is of interest 
for the prediction of variability of the climate. {In fact it has recently been shown 
by Cao and 
Titi \cite{CT07} that the solutions of the viscous HPE remain regular (see also \cite{NJ07}).}
It follows that if extreme events do occur in solutions of HPE, then these must actually be 
smooth at sufficiently small scales. Conversely, HPE dynamics, although now known to be regular, 
may still produce extreme events due to the allowed intense stretching and folding of $\bnabla q$ 
and $\bnabla\theta$ under the dynamics of equations (\ref{froz1}) and (\ref{froz2}). 

\vspace{-2mm}
\section{\sf\large\textbf{Summary of main results for the Euler and Navier-Stokes equations}}

Consider the dimensionless form of the incompressible $3D$ Euler and Navier-Stokes equations
\bel{compeul1a}
\frac{D\bu}{Dt} + \theta\,\bk = Re^{-1}\Delta\bu - \bnabla p\,,\qquad
\frac{D~}{Dt} = \pt + \bu\cdot\bnabla\,,
\ee
where the temperature $\theta(\bx,\,t)$ evolves according to 
\bel{compeul2}
\frac{D\theta}{Dt} = \big(\sigma Re\big)^{-1}\Delta\theta\,.
\ee
Information about $\bnabla\theta$ is needed to determine how $\theta(\bx,\,t)$ might 
accumulate into large local concentrations. 
The traditional approach is to study this question through the dynamics of the 
potential vorticity, defined by ($\bom ={\rm curl}\,\bu$ is the vorticity)
\bel{qdef}
q := \bom\cdot\bnabla\theta\,.
\ee
\begin{center}
\par\vspace{-30mm}
\begin{minipage}[htb]{11cm}
\setlength{\unitlength}{.7cm}
\bc
\begin{picture}(11,11)(0,0)
%
\qbezier[400](1,5)(5,3)(8,3)
\qbezier[400](1,5)(.75,5.5)(.5,6)
\qbezier[400](.5,6)(2,5.5)(3,5)
\qbezier[400](3,5)(6,5.8)(9,6.5)
\qbezier[400](8,3)(7,3.5)(7,4)
\qbezier[400](7,4)(6.5,4.1)(5.3,4.4)
\qbezier[400](1,3)(3,4)(8,5)
\qbezier[400](1,3)(1,3.5)(.5,4)
\qbezier[400](.5,4)(1,4.2)(1.7,4.6)
\qbezier[200](9,6.5)(8,5.8)(8,5)
\thicklines
\qbezier[400](3.45,3.9)(3.2,4.5)(3,5)
\put(3.75,4.2){\vector(-1,2){.4}}
\put(4.1,4.4){\makebox(0,0)[b]{$\bdB$}}
\put(8.6,3.5){\makebox(0,0)[b]{$\theta = \mbox{const}$}}
\put(9.8,5.5){\makebox(0,0)[b]{$q = \mbox{const}$}}
\put(5.7,3.4){\makebox(0,0)[b]{$\bnabla\theta\!\nearrow$}}
\put(7,5.2){\makebox(0,0)[b]{$\bnabla q\,\nwarrow$}}
\end{picture}
\ec
\end{minipage}
\end{center}
\par\vspace{-15mm}\noindent
\textbf{\sf Figure 2\,: \scriptsize For the incompressible Euler equations in three dimensions, 
the vector $\bdB = \bnabla Q(q)\times\bnabla\theta$ is tangent to the curve defined by 
the intersection of the two surfaces $q = \mbox{const}$ and $\theta = \mbox{const}$.}
\par\medskip\noindent
The main results for the incompressible Euler and Navier-Stokes equations are summarized in the following\,:
\begin{theorem}
In the cases below, $q$ and $\theta$ satisfy
\bel{qrho1}
\pt q + \mbox{div}\,\big(q\,\bU\big) = 0\,,\qquad \pt\theta + \bU\cdot\bnabla\theta = 0\,,
\ee
and, with $Q(q)$ as any smooth function of $q$, the divergence-free flux vector
\bel{Bdef}
\bdB = \bnabla Q(q)\times\bnabla \theta \,,
\ee
satisfies the stretching relation
\bel{stretch1}
\pt\bdB - \mbox{curl}\,(\bU\times\bdB) = \bdD\,.
\ee
The divergence-less vector $\bdD$ in (\ref{stretch1}) is given by $\bdD = - \nabla(qQ'\mbox{div}\,\bU)\times\nabla\theta$.
\ben\itemsep -1mm

\item For the incompressible Euler equations, $\bU = \bu$ \& thus $\bdD = 0$. In this case $Dq/Dt = 0$, 
so the intersections of the level sets of $q$ and $\theta$ shown in Figure 1 move together with the 
fluid velocity, $\bu$\,;

\item \textit{For the incompressible Navier Stokes equations $\bU$ is defined as}
\bel{vdef1}
q(\bU - \bu) = - Re^{-1}\left\{\Delta\bu\times\bnabla\theta + 
\sigma^{-1}\bom\Delta\theta\right\}\,,\qquad q \ne 0\,.
\ee
\een
Moreover, for any surface $\bS(\bU)$ moving with the flow $\bU$, one finds
\bel{ex1}
\frac{d}{dt}\int_{\bS(\tiny\bU)} \bdB \cdot d\bS
= \int_{\bS(\tiny\bU)}\bdD\cdot d\bS\,.
\ee
\end{theorem}
\textbf{Remark\,:} The three dimensional incompressible Navier-Stokes equations possess only 
Leray's weak solutions while the Euler equations do not even possess these. Thus the 
manipulations used in deriving (\ref{qrho1})--(\ref{ex1}) should be considered as purely formal.
\par\vspace{1mm}\noindent
\textbf{Sketch proof of (\ref{qrho1})--(\ref{ex1})\,:} The derivation of (\ref{qrho1}) follows 
the same standard manipulations that appear in the elegant classic proof of Ertel's Theorem \cite{Ertel42}, 
namely 
\beq{q1A}
\frac{Dq}{Dt} &=& \left(\frac{D\bom}{Dt} - \bom\cdot\bnabla\bu \right)\cdot\bnabla \theta
+ \bom\cdot\bnabla\left(\frac{D\theta}{Dt}\right)\non\\
&=& \big(Re^{-1}\Delta\bom - \bnabla^{\perp}\theta \big)\cdot\bnabla \theta
+ \bom\cdot\bnabla\left((\sigma Re)^{-1}\Delta\theta\right)
\non\\
&=& \mbox{div}\,\big(Re^{-1}\Delta\bu\times\bnabla \theta 
+ (\sigma Re)^{-1}\bom\Delta\theta \big)
\,,
\eeq
where $\bnabla^{\perp} \theta =\bnabla\theta\times\bk$.
The scalar product $\bnabla^{\perp}\theta\cdot\bnabla \theta=0$ and the rest of the 
terms on the right hand side of (\ref{q1A}) have been regrouped as a divergence. 
On using $\mbox{div}\,\bu =0$ one may define $\bU$ {through the equation}
\beq{q2A}
\pt q &=& -\,  \mbox{div}\,\Big(q\bu 
-   Re^{-1}\Delta\bu\times\bnabla \theta 
-(\sigma Re)^{-1}\bom\Delta\theta  \Big)
=:
-\,  \mbox{div}\,(q\,\bU)\,,
\eeq
{in which case $\mbox{div}\,\bU \ne 0$, and 
\bel{thetaex1}
\left(\pt + \bU\cdot\bnabla\right)\theta = \pt\theta + \left\{
\bu - q^{-1}Re^{-1}\left[\Delta\bu\times\bnabla\theta + \sigma^{-1}\bom\Delta\theta\right]
\right\}\cdot\bnabla\theta =  0\,.
\ee}
The flux $\bcalJ = q\,\bU$ was first introduced by Haynes and McIntyre in the context 
of their `impermeability theorem' \cite{HMc87,HMc90}. There have been objections that 
$\bU$ is not a physical velocity \cite{Dan90,Viudez99}, {which have been 
answered by McIntyre in \cite{Mc90}} but in the context of this paper $\bU$ has been 
employed solely as a notational device. The remarkably simple form of (\ref{stretch1}) 
for the incompressible Euler case, in which $\bU=\bu$ and $\bdD = 0$, {was 
derived first in \cite{KurgTat87,KurgPis00,Kurg02}.} Two versions of the proof of 
(\ref{stretch1}) are given in the Appendix, the first using Lie derivatives and the 
second using conventional vector identities. \hfil $\blacksquare$
\par\medskip
The right hand side of (\ref{stretch1}) occurs because $q$ is not a scalar function\,; rather 
it is a density (a volume form). However, because $\mbox{div}\,\bu = 0$ for the incompressible 
Euler case, it follows that $\bdB$ satisfies 
\beq{stretch2}
\frac{D\bdB}{Dt}  = \bdB\cdot\bnabla\bu
\eeq
which is also the standard stretching equation for vorticity $\bom$ on replacing $\bdB$ with $\bom$. 
The squared magnitude $|\bdB|^2$ satisfies 
\bel{ev1}
\frac12\frac{D~}{Dt}|\bdB|^{2} = \bdB\cdot\sfS\bdB
\approx   \lambda^{\tiny (\sfS)}|\bdB|^{2}\,,
\ee
where $\lambda^{\tiny\sfS}(\bx,\,t)$ is an estimate for an eigenvalue of the rate of strain matrix 
$\sfS$. Alignment of $\bdB$ with a positive (negative) eigenvector of $\sfS$ will produce exponential  
growth (decay), thus mimicking the stretching mechanism that produces the large vorticity intensities 
that develop locally in turbulence. {Ohkitani \cite{Ohk08} has studied Clebsch-decomposed solutions 
for $\bom = \bnabla f\times\bnabla g$ where $Df/Dt =0$ and $Dg/Dt =0$.}
\par\smallskip\noindent
\textbf{Remark\,:} Moffatt suggested the analogy between the magnetic field in a conducting fluid and 
the vorticity in an incompressible Euler flow \cite{HKM1} (see also \cite{Palmer88}). Equation 
(\ref{stretch2}) continues this analogy. Moffatt's detailed discussion of the topology of magnetic 
field lines is based on the concept of helicity that requires the existence of a vector potential 
$\bA$ that satisfies $\bdB = \mbox{curl}\,\bA$ where
\bel{Adef}
\bA = \shalf\big(Q\bnabla\theta - \theta\bnabla Q\big) + \bnabla\psi\,.
\ee
The helicity $H$ that results from this definition, 
\beq{heldef}
H = \int_{V}\bA\cdot\bdB\,dV = \int_{V}\mbox{div}\,\big(\psi\bdB\big)\,dV 
= \oint_{\partial V} \psi \bdB\cdot\bn\,dS\,,
\eeq
measures the winding number, or knottedness of the lines of the divergence-free vector field $\bdB$. 
This helicity would vanish for homogeneous boundary conditions. However, if realistic topographies 
were taken into account then the possibility for $H\neq 0$ would exist. The boundaries may therefore be an important 
generating source for helicity, thus allowing the formation of knots and linkages in the $\bdB$-field.

\section{\sf\large\textbf{The case of the hydrostatic primitive equations}}\label{HPE}

Many simulations of weather, climate and ocean circulation employ the hydrostatic version of the 
primitive equations (denoted HPE). The major difference of HPE 
from the Navier-Stokes equations lies in the exclusion of the vertical velocity component 
$w(x,y,z,t)$ in the hydrostatic velocity field\,\footnote{\sf The primitive equations as used 
for weather and climate prediction are defined on a corrected spherical grid.}
\bel{vdef}
\bv(x,y,z,t) = (u,\,v,\,0)\,.
\ee
However, this vertical component does appear in the transport velocity field $\bV = (u,\,v,\,\varep w)$, 
where $\varepsilon$ is the Rossby number. The velocity field $\bv$ in (\ref{vdef}) obeys the motion equation
\bel{hpe1}
\varep\big(\pt + \bV\cdot\bnabla\big)\bv + \bk\times\bv + a_{0}\bk\Theta 
= \varep\Rey^{-1}\Delta\,\bv - \bnabla p\,,
\ee
and is solved in tandem with the incompressibility condition $\hbox{div}\,\bV = \hbox{div}\,\bv 
+ \varep w_{z}=0$. The vertical velocity $w$ has no evolution equation; it appears only in 
$\bV\cdot\bnabla$  and is determined from the vertical integral of the incompressibility 
condition $\hbox{div}\,\bV = 0$. The $z$-derivative of the pressure field $p$ and the 
dimensionless temperature $\Theta$ enter the problem through the hydrostatic equation 
\bel{azero}
a_{0}\Theta + p_{z} = 0\,,
\ee
which has been been incorporated into (\ref{hpe1}) as its vertical component. The quantity 
$a_{0}$ is a constant 
$\alpha_{a} = H/L\ll 1$ is the aspect ratio and $R_{a}$ is the Rayleigh number.
which comes from the non-dimensionalization of the original equations. By using the 
vector identity 
\bel{hpe2}
\bV\cdot\bnabla\bv = -\bV\times\bzeta +\shalf\bnabla\big(u^{2} + v^{2}\big)
\ee
in (\ref{hpe1}), the vorticity equation for 
\bel{zetadef}
\bzeta = \mbox{curl}\,\bv
\ee
is expressed as
\bel{hpe3} 
\big(\pt + \bV\cdot\bnabla\big)\bzeta = (\sigma\Rey)^{-1}\Delta\bzeta +
\bzeta\cdot\bnabla\bV + \mbox{curl}\bdf\,,
\ee
where $\bdf = -\varep^{-1}\big(\bk\times\bv + a_{0}\bk\Theta\big)$. The dimensionless 
temperature $\Theta$, with a specified heat transport term $h(x,y,z,t)$, satisfies 
\bel{hpetheta}
\big(\pt + \bV\cdot\bnabla\big)\Theta = (\sigma\Rey)^{-1}\Delta\Theta + h\,.
\ee
Equations (\ref{hpe3}) and (\ref{hpetheta}) for HPE correspond to those for the Navier-Stokes 
equations, with the additional $h$-term. Hence the results in \S2 can be lifted over to 
HPE by defining 
\bel{qzdef}
\sfq = \bzeta\cdot\bnabla\Theta\qquad\mbox{and}\qquad
\bsfB = \bnabla \sfQ\times\bnabla\Theta\,,
\ee
where $\sfQ(\sfq)$ can be chosen as any smooth function of the potential vorticity $\sfq$,  which itself obeys 
the relations
\beq{qz1a}
\pt\,\sfq + \mbox{div}\,\big(\sfq\bsfU\big) = 0\,,\qquad\qquad
\sfq\,\big(\pt + \bsfU\cdot\bnabla\big)\Theta = 0\,.
\eeq
Here, the formal transport velocity, $\bsfU$, is defined from the vorticity flux density as
\bel{qz2a}
\sfq\bsfU = \sfq\bV - 
\big\{\big[\Rey^{-1}\Delta\bV + \bdf\big]\times\bnabla\Theta + 
\big[(\sigma\Rey)^{-1}\bzeta\Delta\Theta + h\big]\big\}\,.
\ee
Clearly, $\bsfU$ includes the effects of rotation within $\bdf$ and the heat transport 
term $h$\,: moreover, $\mbox{div}\,\bsfU \neq 0$. As in the case of the Navier-Stokes 
equations this is not a physical velocity but is again a convenient mathematical device. 
Accordingly $\bsfB$ in (\ref{qzdef}) evolves according to the driven stretching relation
\bel{qz2c1}
\pt\bsfB - \mbox{curl}\,(\bsfU\times\bsfB) = \bsfD
\ee
where
\bel{qz2c2}
\bsfD = - \bnabla\Big(\sfq\sfQ'(\sfq)\,\mbox{div}\,\bsfU\Big)\times\bnabla\Theta\,.
\ee
This equation is the analogue for HPE of equation (\ref{qrho1}) for the Navier-Stokes 
case. It implies  that the time rate of change of the flux of $\bsfB$ in (\ref{qzdef}) 
through any surface $\mathbf{S}$ that is  transported at formal velocity $\bsfU$ is given by
\bel{qz2d}
\frac{d}{dt}\int_{\mathbf{S}({\tiny\bsfU})} \bsfB \cdot d\mathbf{S}
= \int_{\mathbf{S}({\tiny\bsfU})}\bsfD
\cdot d \mathbf{S}\,.
\ee
In particular, when the $\bsfU$-transported surface is chosen to be a level set 
of temperature (whose normal vector is along $\bnabla \Theta$ then the right hand side 
vanishes, as it should to maintain the tangency of $\bsfB$ to such surfaces. This means 
that the effects of PV gradient flux creation due to the right hand side of (\ref{qz2c1}) 
occur only on $\bsfU$-transported surfaces that are \emph{not} temperature 
iso-surfaces. 

\section{\sf\large\textbf{Conclusion}}

The equations for the evolution of the flux of PV gradient $\bdB$ for Navier-Stokes in \S2 
and $\bsfB$ for HPE in \S3 are the first two main results. Their left hand sides represent 
the familiar flux transport form that governs the stretching processes while the second 
main feature is the derivation of the right hand side $\bdD$ and $\bsfD$-terms\,; these 
divergence-less forcing terms deserve more investigation. {Herring, Kerr and 
Rotunno \cite{HKR94} have performed a computational study of vortex re-connection in the 
Navier-Stokes Boussinesq system (\ref{compeul1a}) and (\ref{compeul2}). It is possible 
that the divergence-less vector $\bdD$ may be the key to understanding this phenomenon 
with the left hand side of (\ref{froz1}) dominating for early to intermediate times until 
the effect of $\bdD$ destroys the frozen-in property. A numerical study of the effect of
 $\bdD$ may therefore be worthwhile.}

In GFD, topography has been found to have some bearing on the nature of the topology of 
the $\bsfB$-field, because helicity 
is generated at boundaries, and this potentially leads to the formation of knots and 
linkages in the $\bsfB$-field lines. The atmospheric or oceanic events to which these 
knots and linkages would  correspond are not wholly clear.
\par
Investigation of steady-state balances of $\bsfB$ and its interaction with imposed steady coherent 
shear would also be interesting. In fact, analogous investigations of the spatial distribution of 
the PV flux are already underway in other contexts and, for example, have recently been studied 
via the data analysis of PV fronts at the sea surface (Czaja and Hausmann \cite{Czaja08}). We hope 
modern developments in observation and data analysis will soon provide new insight into the role and 
magnitude of the dynamical effects and balance effects caused by the transport of PV gradient flux 
along temperature iso-surfaces. In this regard, see, for instance, the recent paper by McWilliams \textit{et~al} 
\cite{MCM09} for a discussion of filamentary intensification in the ocean by processes similar to 
the stretching of $\bsfB$. Likewise, in the atmosphere, the stretching of $\bsfB$ and the associated alignment properties of $\bnabla q$ and $\bnabla\theta$ are of interest, particularly in the region of the tropopause, as shown in Figure \ref{Flo}.
\rem{The strategy used is directly applicable to compressible flow before the Boussinesq approximation 
is imposed\,: see Spiegel and Veronis \cite{SV60}
\bel{extra1}
\frac{D\bu}{Dt} + g\bk = \bdF -\frac{1}{\rho}\bnabla p\,;\qquad
\frac{D\theta}{Dt} = G\,;\qquad
\frac{D\rho}{Dt} + \rho\,\mbox{div}\,\bu = 0\,, 
\ee
where $\bdF = \rho^{-1}\left\{\nu\Delta\bu + \sthird\nu\bnabla(\mbox{div}\,\bu)\right\}$ and 
$G = G(\theta,\,\rho,\,p,\,\bu)$. Then, provided the pressure satisfies any differentiable 
thermodynamic relation $p = p(\rho,\,\theta)$, the velocity $\bU$ comes out to be
\bel{compUdef}
q(\bU - \bu) = -\left(\bdF\times\bnabla\theta + G\bom\right)\,,
\ee
so the stretching relation (\ref{stretch1}) is still valid.}

Finally, the  fundamental stretching mechanism in either HPE, or Navier-Stokes is the term 
$\bsfB\cdot\bnabla\bsfU$. As already noted, the relation 
$\bsfB\cdot\big(\bsfB\cdot\bnabla\bsfU\big) = \bsfB\cdot\sfS\bsfB$ implies that alignment 
of $\bsfB$ along a positive eigenvector of 
the rate of strain matrix $\sfS$ (noting that $\mbox{div}\,\bsfU \neq 0$) will lead to 
exponential growth in $\bsfB$. Thus, there will be a tendency for $\bsfB$ to stretch in 
these positive directions within a large coherent vortex.  It has long been observed 
that large-scale vortices develop plateaus in PV and form steep cliff-like edges at the 
vortex boundary\,: see Rhines and Young \cite{RhinesYoung82a} and Rhines 
\cite{Rhines93}. As noted above, although the effect of the $\bdD$-term in (\ref{qz2c1}) 
is not yet clear, it may play a significant role in the mechanism by which plateaus are 
formed in PV profiles; namely, by the transport of PV gradient within a region of nonzero 
PV and along temperature iso-surfaces. The fact that PV gradient flux can be  created along 
temperature iso-surfaces and can penetrate any other $\bsfU$-transported surfaces may also 
help explain the `leakage', or `erosion' of PV gradient that is observed in certain regions 
of vortex boundaries.
\par\medskip\noindent
\textbf{Acknowledgements\,:} We are grateful to P. Berloff, M. Bustamante, C. J. Cotter, R. Hide, B. Hoskins, N. Klingaman. P. Lynch \& J. T. Stuart for discussions. DDH also thanks the 
Royal Society of London Wolfson Scheme for partial support.

\vspace{-4mm}
\appendix
\section{\sf\textbf{\large Derivation of equation (\ref{stretch1})}}

\noindent
Given the advective transport equations for temperature and potential vorticity,
\beq{a1}
\frac{D\theta}{Dt} = 0\,,\qquad\qquad
\frac{D}{Dt}\big(q\,d^{3}x\big) = \big(\pt q + \bu\cdot\bnabla q + 
q\,\mbox{div}\,\bu\big)\,d^{3}x = 0\,,
\eeq
the evolution equation (\ref{stretch1}) for the quantity $\bdB = \bnabla Q(q)
\times\bnabla\theta$ may be derived easily by using the notation of the exterior 
derivative $(\,d\,)$ and the wedge product $(\,\wedge\,)$
\bel{a4}
\bdB\cdot d\mathbf{S} = \big(\bnabla Q(q) \times\bnabla\theta\big)\cdot d\mathbf{S}
 = dQ(q)\wedge d\theta\,.
\ee
The advective time derivative of the leftmost term in this relation yields
\bel{a5}
\frac{D~}{Dt}\big(\bdB\cdot d\mathbf{S}\big) 
= \left[\pt\bdB - \mbox{curl}\,(\bu\times\bdB) \right]\cdot d\mathbf{S}
\qquad\hbox{along}\qquad \frac{D\bx}{Dt} = \bu
\ee
The advective time derivative of the rightmost term in (\ref{a4}) 
yields, using equations (\ref{a1}),
\begin{eqnarray}\label{comp-a}
\frac{D~}{Dt}\Big(dQ(q)\wedge d\theta \Big) &=& d\left(\frac{DQ(q)}{Dt}\wedge d\theta\right) 
+ dQ(q)\wedge d\left(\frac{D \theta}{Dt} \right)\non\\
&=& -\,d\left(qQ'\,\mbox{div}\,\bu\right)\wedge d\theta 
= \bdD\cdot d\mathbf{S}\,,
\end{eqnarray}
also along ${D\bx}/{Dt} = \bu$. For incompressible Euler flow, $\mbox{div}\,\bu = 0$ 
and equation (\ref{stretch1}) arises by equating the rightmost terms in (\ref{a5}) 
and (\ref{comp-a}). When $\bu$ is replaced by $\bU$ for the Navier-Stokes equations, 
this yields a non-zero right hand side because in this case $\mbox{div}\,\bU \neq 0$. 
\par\smallskip\noindent
The second version of the proof, with the notation $\bom_{U}= \mbox{curl}\,\bU$, is simply a direct calculation\,: 
\beq{stretch3}
\bdB_{t} &=& (\nabla Q)_{t}\times(\nabla\theta) + (\nabla Q)\times(\nabla\theta)_{t}\non\\
&=& -\nabla\big[(qQ'\,\mbox{div}\,\bU) + \bU\cdot\nabla Q)\big]\times(\nabla\theta)
- (\nabla Q)\times\left[\nabla(\bU\cdot\nabla\theta)\right]\non\\
&=& -\left\{\nabla(qQ'\,\mbox{div}\,\bU) + \bU\cdot\nabla(\nabla Q) + (\nabla Q)\cdot\nabla\bU
+ (\nabla Q)\times\bom_{U}\right\}\times(\nabla\theta)\non\\
&\quad& -\,(\nabla Q)\times\left\{\bU\cdot\nabla(\nabla\theta) + 
(\nabla\theta)\cdot\nabla\bU + (\nabla\theta)\times\bom_{U}\right\}\non\\
&=& -\nabla(qQ'\,\mbox{div}\,\bU)\times\nabla\theta - \bU\cdot\nabla\bdB 
+ (\nabla Q)(\bom_{U}\cdot\nabla\theta) - (\nabla\theta)(\bom_{U}\cdot\nabla Q)\non\\
&\qquad&  +\,(\nabla\theta)\times(\nabla Q\cdot\nabla\bU) - (\nabla Q)\times(\nabla\theta\cdot\nabla\bU)\non\\
&=& \mbox{curl}\,(\bu\times\bdB) -\nabla(qQ'\,\mbox{div}\,\bU)\times\nabla\theta\,.
\eeq

\par\vspace{-2mm}


\end{document}